\documentclass[prb]{revtex4}
\usepackage{graphicx}
\usepackage{dcolumn}
\usepackage{bm}
\usepackage{color}

\begin{document}

\preprint{IFUNAM FT-032}

\title{Simulation of an inhomogeneous Fermi gas through the BCS-BEC crossover}

\author{R. J\'auregui, R. Paredes, L. Rosales-Z\'arate,
and G. Toledo S\'anchez}

\affiliation{ Departamento de  F\'{\i}sica Te\'orica, Instituto de
F\'{\i}sica, Universidad Nacional Aut\'onoma de M\'exico, A.P. 20-364,
M\'exico 01000 D.F. M\'exico}

\date{\today}

 \begin{abstract}
 We perform a variational quantum Monte Carlo simulation of
 the transition from a Bardeen-Cooper-Schrieffer superfluid (BCS)
 to a Bose-Einstein condensate (BEC) at zero temperature.
 The model Hamiltonian involves an attractive short range two body
 interaction and the atoms number $2N =330$ is chosen so that,
 in the non-interacting limit, the ground state function  corresponds
 to a closed shell configuration. The system is then characterized
 by the $s$-wave scattering length $a$ of the two-particle collisions in the gas,
which is varied from negative to positive values, and the Fermi wave
number $k_F$. Based on an extensive analysis of the $s$-wave
two-body problem, one parameter variational many-body wave functions
are proposed to describe the ground state of the interacting Fermi
gas from BCS to BEC states.  We exploit properties of
antisymmetrized many-body functions to develop efficient techniques
that permit variational calculations for a large number of
particles. It is shown that a virial relation between the energy per
particle and the trapping energy is approximately valid for
$-0.1<1/k_Fa<3.4$. The influence of the harmonic trap and the
interaction potential as exhibited in two-body correlation functions
is also analyzed.

\end{abstract}

 \maketitle

\section{INTRODUCTION}
The experimental realization of a degenerate Fermi gas in 1999
\cite{D.Jin}, boosted  theoretical and experimental efforts to
study interacting Fermi gases, in particular, the formation of
molecules and highly correlated pairs from a balanced mixture of
neutral interacting Fermi atoms in two different hyperfine spin
states \cite{experiments, thomas, bartenstein, salomon, bourdel,
heiselberg}. The possibility of tuning the strength of the
interaction between particles in different spin states via
Feshbach resonances, results in the formation of Cooper pairs
(molecules)  for negative (positive) values of the scattering
length $a$. At low temperatures, these pairs  and molecules  can
form a Bardeen-Cooper-Schrieffer (BCS) superfluid state and a
Bose-Einstein condensate (BEC) respectively. When crossing from
the BCS to the BEC region, and viceversa, $a$ grows in magnitude
until it diverges at the resonance. In this limit the scattering
length is no longer a relevant scale, and the properties of the
gas become independent of the specific details of the interaction
potential. This is the so called unitarity limit in which the gas
is assumed to be universal\cite{thomas, heiselberg,
ho,wernercastin}, because its properties depend locally just on
the density and the temperature, i.e.,  the only relevant scales
in this quantum gas are the interparticle spacing and the Fermi
energy. Consequently the gas properties can be expressed in terms
of them and universal parameters \cite{thomas, heiselberg, ho}.

Previous treatments of the BCS-BEC crossover in degenerate atomic
gases have been done using different approaches, we can mention
the self-consistent many-body approach \cite{heiselberg}, the
effective field theory \cite{field-theory}, and more recently
quantum Monte Carlo calculations \cite{carlson,pandharipandepra,
giorgini, Dlee, chang07, BlumeD07, BlumeD08}. This last treatment
has been predominantly based on the fixed node Quantum Monte Carlo
technique. In most of these calculations, the two-component Fermi
gas is considered as an homogeneous system although,
experimentally, the gas has an intrinsic inhomogeneous nature
provided usually by a magnetic and/or optical trap. Such confining
can be described by a harmonic potential. An interesting parameter
calculated in those approaches is $\beta$, which relates the Fermi
energy of the ideal Fermi gas $E_{IFG}$ and the total energy of
the interacting gas $E$. This parameter is expected to acquire a
universal value at unitarity \cite{ho}. The predicted values for
$\beta$ ranges  from -0.75 to -0.33
\cite{heiselberg,field-theory,carlson,giorgini,Dlee}. First
experimental estimates gave $\beta \sim - 0.36$ \cite{salomon} and
$\beta\sim -0.49\pm 0.04$ \cite{kinast} while, more recently,
values around $-0.54$\cite{hulet,regal} have been reported. The
later results are based on measurements of the gas cloud radii at
unitarity.

In a recent work \cite{rapidA}, we employed variational quantum
Monte Carlo techniques (VQMC)  to describe a balanced
two-component interacting gas confined in a three-dimensional
harmonic potential. There, we reported direct
  tests of the universality hypothesis in the unitarity limit that include:
(i) the verification of virial relations for  $N$= 4, 10, 20, 35,
56, 84, 120, 165  and 220, (ii) the variational
 estimate $\beta_{fit}\ge -0.50^{(+0.02)}_{(-0.04)}$
  using a linear fit of the energy per particle.
   In that paper we also briefly reported an analysis on observables like the
system energy and density profiles in the BCS-BEC crossover. In
particular we found $N-$independent energy curve features through
the crossover.

 In the present article, we extend the
analysis  of  Ref.~\cite{rapidA}, paying special attention to
exhibit additional properties of the trial many-body wave functions,
whose structure incorporates, in an analytic and compact form,
important features of the trapped two-body system. Particular
properties of the antisymmetrized many-body functions, let us
develop efficient techniques that permit variational calculations
for an unusual large number of particles. The optimized wave
functions allow the study of the influence of the harmonic trap and
the interaction potential in energies, densities and two-body
correlation functions, all along the crossover. The correlations
between atoms in the same hyperfine state show the Pauli-blocking
evolution as a function of $a$. Similarly, the correlations for
atoms in different hyperfine states give information on the
formation of molecules and Cooper pairs. The applicability of virial
relations to our results is also analyzed. Here we report results
for $N=165$ particles per each hyperfine state.

 This work is organized as follows: in section II, an
extensive discussion of the two-body problem in the trap is done,
and expressions for two-body functions that contain interaction
and trap effects are obtained. The results of that section are
then used to construct variational many-body wave functions for
each region of the crossover. In section III, we address the
many-body system and exploit the structure of the variational wave
functions to optimize numerical calculations. There, we describe
in detail the procedure for the variational quantum Monte Carlo
simulation and energy evaluation. This section also contains the
results for optimal variational parameters and energies, as well
as densities and two-body correlation profiles. Our conclusions
are presented in section IV.

\section{THE TWO-BODY PROBLEM}

In this section we shall establish the two-particle system
features.
 As it is well known, in the limit of low energies it is expected
that the scattering process, represented by the $s$-wave
scattering length $a$, determines the general features of the
state of two colliding particles, regardless the detailed form of
the interaction potential among them.

Here we consider two particles of mass $m$ trapped in a harmonic
potential of frequency $\omega$ and interacting through an
isotropic attractive potential of finite range $b/2$ given by
\begin{equation}
V(r_{i,j})=V_0e^{-2\vert {\bf r}_{i\uparrow} - {\bf
r}_{j\downarrow}\vert/b}, \quad V_0 < 0 \label{twobody}
\end{equation}
where the $\uparrow$ and $\downarrow$ subindices denote two
different hyperfine atomic states and $b<<\sqrt{\hbar/m\omega}$. The
potential is chosen so that, in otherwise free space, it would
admit a finite number of bound states as its strength $V_0$ is varied.\\

For interactions taking place in free space, the Schr\"odinger
equation
 \begin{equation}
[\frac{{ p}^2}{m} + V]\phi = {\cal E}\phi \label{eq:schr}
 \end{equation}
 has analytical $s$-wave solutions \cite{rarita} $\phi(r) =
v(r)/r$ both in the continuum
\begin{equation}
v(y) = c_1J_{i b\sqrt{{\cal E} m}/\hbar}(y) +c_2J_{-ib\sqrt{{\cal E}
m}/\hbar}(y),
\end{equation} and in the bound states region
\begin{equation}
 v(y) =c_+J_{ b\sqrt{\vert {\cal E}\vert m}/\hbar}(y)
 \label{eq:bess}
 \end{equation}
where $y = \zeta e^{-r/b}$, $\zeta=(b\sqrt{\vert V_0\vert
m}/\hbar)$, and $J_\nu$ represents the Bessel function of the first
kind of order $\nu$. By imposing the proper boundary conditions and
considering the limit ${\cal E}\rightarrow 0^+$, the following
expression is found for the $s$-wave scattering length dependent
just on $\zeta$
\begin{equation} a = -b\Big[ \frac{\pi}{2} \frac{N_0(\zeta)}
{J_0(\zeta)} - \log( \zeta/2) -C \Big],\label{scatteringlength}
\end{equation}
with $N_0$ the Bessel function of the second kind and order zero,
and $C$ the Euler constant. This scattering length diverges whenever
$J_0(\zeta)=0$. Denoting the zeros of the $J_0$ Bessel function in
increasing order by $z_k$ $(k=0,1,2,...)$, the potential $V(r)$
admits just $k$-bound states for $z_k< \zeta < z_{k+1}$. The
discrete eigenvalues are determined by the boundary condition at
$r=0$, $J_{b\sqrt{\vert {\cal E}\vert m}/\hbar}(b\sqrt{\vert
V_0\vert m}/\hbar)=0$.

When the two-body collision process takes place in the presence of
an isotropic harmonic potential, the two-body Schr\"odinger
equation can be separated in a center of mass equation
\begin{equation}
[\frac{ {{ P}^2_{CM}}}{2M } +\frac{1}{2} M \omega^2{{
R}^2_{CM}}]\Phi({{\bf R}_{CM}}) = E_{CM}\Phi({{\bf R}_{CM}}),
\end{equation}
 and a relative coordinate equation
\begin{equation}
[\frac{ { p}^2}{2\mu } + \frac{ 1}{2}\mu \omega^2{
r}^2+V(r)]\varphi({\bf r}) = \epsilon\varphi({\bf r}).
\label{eq:rel}\end{equation} with $\mu=m/2$ and $M=2m$. The former
is the isotropic harmonic oscillator equation whose solutions are
well known, and the latter can be numerically solved  given $b$ and
$V_0$.

\begin{figure}
\includegraphics[width=3in]{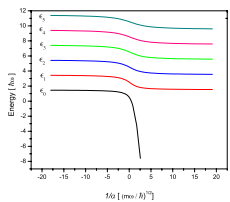}
\caption{(Color online) Lowest $s$-wave relative energy
eigenvalues in units of $\hbar\omega$  for  two colliding trapped
particles, Eq.~(\ref{eq:rel}), around the first resonance. It was
evaluated by considering a  potential range
$b/2=0.015\sqrt{\hbar/m\omega}$ and a strength $V_0$ starting from
$V_0\sim 0$ to the lowest $\vert V_0\vert$ yielding $a\rightarrow
0^+$. The scattering length is measured in units of
$\sqrt{\hbar/m\omega}$.} \label{fig_1}
\end{figure}

Figure \ref{fig_1}  illustrates the $s$-wave lowest eigenvalues
$\epsilon_n$, $n\le 5$, as a function of the inverse of the
scattering length when the potential parameters are in the first
resonance region ( $\zeta =b\sqrt{\vert V_0\vert m}/\hbar$ around
$z_0$) and $b<<\sqrt{\hbar/m\omega}$. For $- \infty <1/ a < 0$, the
spectrum is discrete with positive values in counterpart to the free
space system which has a continuum spectrum and no bound states. In
fact for $a\rightarrow 0^-$, $\epsilon_n\cong (2n+3/2)\hbar\omega$
as expected. At resonance, $1/a=0$, the ground state energy
$\epsilon_0\sim 1/2 \hbar \omega$. For positive $a$,  $\epsilon_0$
decreases becoming zero at $1/a\sim 1/2\sqrt{\hbar/m\omega}$. For
$a\rightarrow 0^+$, $\epsilon_0$ takes values close to the ground
state energy of the free space system, Eq.~(\ref{eq:schr}), while
the excited states energies become $\epsilon_{n>0}\cong
(2n-1/2)\hbar\omega$. In the whole region $-\infty < 1/a < \infty$,
all $\epsilon_{n>0}$ exhibit a similar behavior and are
consecutively spaced among them by  a factor of $\sim 2\hbar\omega$.
In fact, at resonance, $\epsilon_n\cong (2n + 1/2) \hbar \omega$.

If $\zeta$  is further increased, the scattering length becomes
negative  until the second resonance is reached at $z_1$. Around the
second resonance, the eigenvalue $\epsilon_{t+1}$ as a function of
$a$ is similar to $\epsilon_t$ around the first resonance. For
instance, for $\zeta =z_1$, the $first$ excited state energy
$\epsilon_1$ becomes $\epsilon_1 \cong \hbar\omega/2$. Meanwhile,
the ground state energy $\epsilon_0$ remains similar to the
corresponding ground state energy of Eq.~(\ref{eq:schr}) which
decreases with growing $\zeta$. Higher values of $\zeta$ yield
analogous results, so that $\epsilon_{k+t}$, $t=0,1,2,...$ as a
function of $1/a$ around the $(k+1)^{th}$-resonance is similar to
$\epsilon_t$ around the first resonance.

As expected, the qualitative behavior of the spectrum illustrated in
Fig.~1  using the  finite range interaction $V$, is in excellent
agreement with the analytical results for two harmonically trapped
particles interacting through a regularized contact potential
$(4\pi\hbar^2 a/m)\delta_{reg}({\bf r}_i - {\bf r}_j)$
\cite{wilkens}. For that problem, Busch $et$ $al$ found an implicit
equation for the energy eigenvalues $\epsilon$,
\begin{equation}
\sqrt{2}\frac{\Gamma(-\epsilon/(2\hbar\omega)
+3/4)}{\Gamma(-\epsilon/(2\hbar\omega) +
1/4)}=\frac{\sqrt{\hbar/m\omega}}{a} \label{eq:wilkens},
\end{equation}
and the explicit expression for the corresponding eigenfunctions. In
particular, for $\vert a\vert\rightarrow\infty$, the ground state
energy is $\epsilon_0 = \hbar\omega/2$. We have checked that given
$a$, the finite range interaction spectrum $\epsilon_{k+t}$,
$t=0,1,2,...$ around the $(k+1)^{th}$-resonance  reproduces with
increasing accuracy the contact interaction spectrum as the
potential range parameter $b\rightarrow 0$. In order to obtain such
 matching,  shorter potential ranges $b/2$ are required  for
negative energies than for positive energies.

From now on, we consider just  short range potentials and $\zeta$
around the first resonance condition. The general behavior of the
$s$-ground state eigenfunctions $\varphi_{a}(r)$ is illustrated in
Fig.~2 and Fig.~3 in terms of  the functions $u_{a}$
($\varphi_{a}(r)=u_{a}(r)/r$)  considering the free space and
trapped system respectively. In both figures, the solid line
represents $u_\infty$ at resonance ($a = \infty$). The structure of
this ground state trapped wave function deviates significantly from
its free-space analog not just at long distances both also near the
origin.

For $a<0$, we have found that the numerical solution can be
approximated using the following analytical compact
representation:
\begin{equation}
\varphi_{apx}(r) = J_0(z_0 e^{-r/b})e^{-m\omega r^2/4\hbar}(1 +c
e^{-2r/b})P(r/b)/r \label{eq:form1}
\end{equation}
where $c$ is independent of $r$ and $P(r/b)$ is a polynomial
function. In fact, this approximation  has an accuracy higher than
$0.01\%$ by the proper choice of $c$ and a fourth order polynomial
$P(r/b)$, both of which depend on $V_0$ and $b$. The accuracy of
this approximation was measured by evaluating the ratio
$\varphi_{apx}(r)/ \varphi_{num}(r)$ between the analytical
approximate expression Eq.~(\ref{eq:form1}) and the numerical
solution.

\begin{figure}
\includegraphics[width=3in]{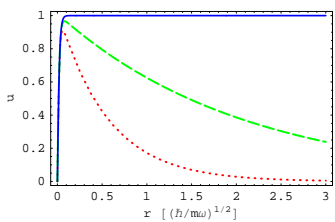}
\caption{(Color online) Radial function $u_a(r)$  for interacting
particles in otherwise free space. The zero-energy resonant
function $u_\infty(r)$ (solid line) tends to a nonzero constant as
$r\rightarrow \infty$, meanwhile $u_{2.1}(r)$ (dashed line) and
$u_{0.58}(r)$ (dotted line) correspond to increasingly bound
states. Distances are measured in units of
$\sqrt{\hbar/m\omega}$.} \label{fig_2}
\end{figure}

\begin{figure}
\includegraphics[width=3in]{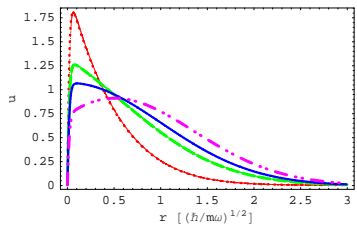}
\caption{(Color online) Radial function $u_a(r)$ for interacting
particles in the presence of the trapping potential. The
dot-dashed curve corresponds to the ground $s$-state for a
negative scattering length $u_{-0.6}(r)$, the resonant function
$u_\infty$(r) is given by the solid curve, while $u_{2.1}(r)$ by
the dashed one and $u_{0.58}(r)$ by the dotted line. In this
figure the wave functions have been properly normalized. Distances
are measured in units of $\sqrt{\hbar/m\omega}$.} \label{fig_3}
\end{figure}

For $\zeta>z_0$  in the region of positive $a$, the {\it ansatz}
for the ground state function is:
\begin{equation}
\varphi_{apx}(r) = v(y(r))e^{-m\omega r^2/4\hbar}g(r)/r
\label{eq:form2}
\end{equation}
where $v$ was defined  in Eq.~(\ref{eq:bess}). The function
$\varphi_{apx}$ is numerically accurate at least at the $1\%$ level.
Its structure let us understand the origin of the eigenvalue
$\epsilon\sim1/2\hbar\omega$. In this case, $v(y(r))$ takes care of
the boundary condition $v(0) = 0$ so that the effective equation for
$g(r)$ is almost identical to its analog for the one dimensional
harmonic oscillator without the requirement of becoming null at $r =
0$, thus admitting the possibility $\epsilon=\hbar\omega/2$.

 The analytical approximations given by Eqs(\ref{eq:form1}-\ref{eq:form2})
  to the exact solutions of the two-body problem will
be exploited in the study of the many-body system.

\section{THE MANY-BODY SYSTEM}
Let us consider the system made up of $2N$ fermions of mass $m$ in
two, equally populated, hyperfine states ($N=N_{\uparrow}
=N_{\downarrow}$) confined in an isotropic three-dimensional
harmonic trap of frequency $\omega$. The system is allowed to
interact via collisions between particles of different hyperfine
states. The Fermi gas is considered to be at zero temperature and
the two-body collision process is approximated by the single-channel
model described in the previous section, Eq.~(\ref{twobody}). The
Hamiltonian describing such a system is:
\begin{eqnarray}
H&=&H_{trap} + \sum_{i,j} V(r_{i,j})\nonumber\\
&=&\sum _{i,j=1} ^{N}\Big[\frac{p_{i\uparrow}^ {2}
+p_{j\downarrow}^ 2}{2m}+
  \frac{1}{2}m\omega^2 \left( r_{i\uparrow}^2 + r_{j\downarrow}^2\right)\Big]+
\sum_{i,j=1}^N V(\vert {\bf r}_{i\uparrow}- {\bf r}_{j\downarrow}\vert)\\
 \label{HMB}
\end{eqnarray}

\subsection{Variational Monte-Carlo simulations}

 In a variational calculation, for a given form of the interaction potential, the
 optimal value of any variational parameter $\lambda$
 in the wave function $\Psi_{\lambda}$, is determined by imposing
  that the expectation value of the  Hamiltonian, Eq.~(\ref{HMB}) in our
problem, to be a minimum with respect  to such parameter. So that,
\begin{equation}
  \frac{\partial E(\lambda)}{\partial\lambda}=0,
  \;\; {\rm where} \;\;
  E(\lambda) = \frac{\langle\Psi_{\lambda}|H|\Psi_{\lambda}\rangle}
  {\langle\Psi_{\lambda}|\Psi_{\lambda}\rangle} \;.
 \label{ELambda1}
\end{equation}
For a system of $2N$~atoms, computing the expectation value requires
the evaluation of a $6N$-dimensional integral. The main idea of the
Monte Carlo method~\cite{Ko86} is not to evaluate the integrand at
every one of the quadrature points, but rather at only a relatively
small representative sampling, where the sequence of configurations
are distributed according to $\vert\Psi_{\lambda}\vert^{2}/
{\langle\Psi_{\lambda}|\Psi_{\lambda}\rangle}$ . We use Metropolis
algorithm~\cite{Me53} which ensures that the desired probability
distribution is approached asymptotically.

 In this article, the
generic form of the variational wave function will be different
according to the region of the BCS-BEC crossover. Explicit details
will be given for each region separately.

\subsubsection{Variational calculation for weakly interacting fermions}

First, let us consider  the region in the potential parameters space
where, for a given range $b/2$, the amplitude of the potential is so
small that no bound states are allowed in the homogeneous two-body
problem. There, it is expected that the trapped ideal Fermi gas
configuration gives a rough description of the system. Accordingly,
a Jastrow-Slater wave function of the form
\begin{equation}
\Psi_{\lambda}^{JS} =\Phi_{IFG}\cdot
F^{J}_{\lambda}\label{eq:Jastrow}
\end{equation}
is assumed. Here $\lambda$ is a variational parameter, $\Phi_{IFG}$
is the Fermi gas wave function given by the product of Slater
determinants (one for each hyperfine state)  describing a
noninteracting system of harmonically trapped atoms, and the Jastrow
function $F^J_\lambda$ will explicitly include the effects of the
interaction potential. It is expected that this kind of wave
function gives an appropriate description of the weakly interacting
Fermi gas in the normal regime but not in the superfluid one.

The inputs of the Slater determinants are the single-particle
eigenstates of a non-interacting particle in a harmonic trap,
$\phi^{ho}_{\bf n}({\bf r})$, with quantum numbers ${\bf n}$ at the
position ${\bf r}$. This construction ensures that the wave function
is totally antisymmetric under the exchange of identical atoms. The
energy of each single particle state is characterized by three
integer quantum numbers ${\bf n}\!\equiv\!(n_x, n_y, n_z)$:
\begin{equation}
  E_{\bf n} = \hbar\omega(\frac{3}{2}+n_x+n_y+n_z) \;,
  \quad (n_{i}=0, 1, 2,   \ldots) \;.
  \label{SPenergy}
\end{equation}
Writing $n=n_x+n_y+n_z$, the degeneracy of each energy level is
$(n+1)(n+2)/2$. A typical basis state has the form:
\begin{equation}
  \phi^{ho}_{n_x,n_y,n_z}({\bf r}) = \left(\frac{1}{a_{ho}^2 \pi}\right)^{3/4}
 \prod_{\xi=x,y,z}
 \frac{ H_{n_i}( \xi/a_{ho})}{\sqrt{2^{n_\xi}n_\xi!}}
 e^{-\xi^2/2a_{ho}^2},
 \label{typical}
\end{equation}
where $ H_{n_\xi}( \xi/a_{ho})$ are the Hermite functions of order
$n_\xi$ and $a_{ho}=\sqrt{\hbar/m\omega}$. In this paper, we
consider closed shell configurations so that the ground state is
built up by taking all single-particle states with energies
increasing from $E_{\bf 0} = 3\hbar\omega/2$ up to the Fermi energy
$E_F=({\cal M}_F + 3/2)\hbar\omega$, where ${\cal M}_F$ is the
maximum energy level for a given number of particles. For large $N$,
$E_F\sim(6N)^{1/3}\hbar\omega$ and the corresponding radius is
$R_F^2=2E_F/m\omega^2$.

In the literature of interacting bosons and fermions, the Jastrow
wave function usually takes the form of a product $\prod_{i,j}
f_{ij}$ of correlation functions $f$ that depend on the degrees of
freedom of the pair $i$, $j$ of interacting particles.  In
Refs.~\cite{carlson,Daniel, bethe73, pandharipande77} $f$ is a
function of the interparticle distance $r$ that solves the
free-space interacting two-body problem up to a healing distance
$d$ after which it is restricted to become constant. In those
works, the parameter $d$ is chosen by minimizing the energy.

In this paper, we shall consider trial many-body wave functions
which yield a continuous $F^J_\lambda$ and continuous derivatives;
the optimal variational parameter $\lambda$ of the trial wave
function for the many-body system will establish an effective $d$
as we illustrate below. In fact, we have studied two options for
the Jastrow function:

(i)$f_{ij}= \exp(-\lambda_{J1} V_0 e^{-2r_{i,j}/b})$, so that,
\begin{equation}
F^J_{\lambda_{J1}}=\exp [-\lambda_{J1}
\sum_{i_\uparrow,j_\downarrow} V(\vert {\bf r}_{i\uparrow}-{\bf r}_
{j\downarrow}\vert )] \label{Psivar}\end{equation}

(ii)\begin{equation}\quad f_{ij} = J_0(z_0
e^{-r_{i,j}/\lambda_{J2}})(1 +c
 e^{-2r_{i,j}/\lambda_{J2}})P(r_{i,j}/\lambda_{J2})/r_{i,j}\label{eq:J2}\end{equation}

The first choice of the variational wave function (\ref{Psivar})
has the advantage of becoming exact when no interactions between
hyperfine states are allowed ($\lambda_{J1}=0$) which is the
trapped ideal Fermi gas limit, where the only correlations are
those imposed by the Pauli exclusion principle. It is inspired on
previous calculations for the nuclear matter \cite{nuclear,
toledoPRC02}, where an appropriate choice of the potential allows
to explore dynamically the interplay of the nuclear-to-quark
matter regime. In addition, this form of the variational wave
function allows to estimate the energy expectation value by
computing only spatial dependent functions in a Monte Carlo
simulation \cite{ceperly}, with no need of calculating the spatial
derivatives of the trial wave function, as we show below.

  The second choice  is inspired on the general structure of the
  two-body wave functions in free space at low energies, Eq.~(\ref{eq:form1}). It allows to
  numerically explore shorter potential ranges than the first option
  (\ref{Psivar}). It reproduces the fact, first noticed in
  BCS theories, that even the slightest interaction can lead to two-body
  long-range-correlations, implicit in the polynomial $P(r_{i,j}/b)$.
  Besides, it increases the reproducibility of interaction effects at short
  interparticle separations through the factor proportional to $c$.
   Deviations from $\lambda_{J2}=b$ should be interpreted as a many-body effect.

 The structure of the variational wave function for the BCS region,
 allows to simplify the expectation value of  the kinetic energy
 operator through an  integration by parts \cite{nuclear}, so that
\begin{equation}
  \frac{\langle\Psi_{\lambda_J}|H_{trap}|\Psi_{\lambda_J}\rangle}
  {\langle\Psi_{\lambda_J}|\Psi_{\lambda_J}\rangle} =
  E_{IFG}
  + 2\sum_{i=1}^{N}\sum^N_{j,j^\prime=1}\frac{\hbar^2}{2m}
  \frac{\langle\Psi_{\lambda_J}\vert{\bf \nabla}_i (\log f_{ij}) \cdot {\bf \nabla}_i (\log f_{ij^\prime})|\Psi_{\lambda_J}\rangle}
  {\langle\Psi_{\lambda_J}|\Psi_{\lambda_J}\rangle},
 \label{Tlambda}
\end{equation}
where $E_{IFG}$ is the  energy of the $2N$ non-interacting trapped
Fermi atoms. For closed shell configurations, this energy can be computed using
the following equation \cite{PRA04}

\begin{equation}
\frac{E_{IFG}}{2N}= \frac{3}{4}\Big[{\cal M}_F + 2\Big]\hbar\omega,
\label{dtfg}
\end{equation}
instead of the large $N$ limit, $E_{IFG}/2N=3{\cal
M}_F\hbar\omega/4$, which produces a slightly underestimated
value. Eq.~(\ref{dtfg}) is valid in general. The extra term in
Eq.~(\ref{Tlambda}) reflects the increase in the kinetic energy of
the system, relative to the Fermi-gas estimate, due to
interactions.

 In the case of Eq.~(\ref{Psivar}), we define the
factors ${\cal W}_{\lambda_{J1}}$ and ${\cal V}_{\lambda_{J1}}$
through the equations
\begin{eqnarray}
    \sum_{i=1}^{N}\sum_{j,j^\prime=1}^N\frac{\hbar^2}{2m}\langle\Psi_{\lambda_{J1}}|
    {\bf \nabla}_i (\log
  f_{ij})\cdot {\bf \nabla}_i (\log
  f_{ij^\prime})|\Psi_{\lambda_{J1}}\rangle &=&
  \frac{\lambda_{J1}^2\hbar^2}{2m}\sum_{i=1}^N
    \langle\Psi_{\lambda_{J1}}|\sum_{j,j^\prime=1}^{N}{\bf \nabla}_i V(r_{i,j})\cdot
    {\bf \nabla}_i V(r_{i,j^\prime})|\Psi_{\lambda_{J1}}\rangle \nonumber\\
    &\equiv& \lambda_{J1}^2 {\cal W}_{\lambda_{J1}}
    {\langle\Psi_{\lambda_{J1}}|\Psi_{\lambda_{J1}}\rangle}, \label{Wpotent}
\end{eqnarray}
and
\begin{equation}
   \sum_{ij} \langle \Psi_{\lambda_{J1}}| V(r_{i,j})|\Psi_{\lambda_{J1}} \rangle \equiv {\cal V}_{\lambda_{J1}}
   {\langle\Psi_{\lambda_{J1}}|\Psi_{\lambda_{J1}}\rangle},
 \label{Vpotent}
 \end{equation}
so that the expectation value of the total energy is:
\begin{equation}
  E(\lambda_{J1}) = E_{IFG} + 2\lambda_{J1}^2 {\cal W}_{\lambda_{J1}}+ {\cal V}_{\lambda_{J1}}.
 \label{ELambda2}
\end{equation}
 The two functions that
remain to be evaluated (${\cal W}_{\lambda_{J1}}$ and ${\cal
V}_{\lambda_{J1}}$) are local; their expectation values may be
computed via Monte Carlo techniques as described above. A similar
approach can be used in the case corresponding to the Jastrow
function Eq.~(\ref{eq:J2}).

We have performed calculations of the energy for a fixed value of
$N$, the range $b/2$ and the scattering length $a$, exploring for
several values of the variational parameter, picking up the one
which minimizes the energy. Each run used about $10^3$ steps for
thermalization and about  $10^4$ more to take data.
  In the first rows of Table I, we report the
 numerical optimal energies using the first choice for the Jastrow
 function and a potential range $b/2 = 0.015\sqrt{\hbar/m\omega}$.
 Similar results are obtained when the second choice of $F^J$ is used.
 The data corresponds to $N=165$, which fills eight shells (${\cal M}_F$ =8)
 for the harmonic potential in three dimensions; $k_F$ represents
 the Fermi wave number associated to the ideal Fermi energy
 $E_{F}= (\hbar k_F)^2/2m$. For $N=165$
the corresponding energy per particle for an ideal Fermi gas is
$E_{IFG}/2N= 7.5 \hbar\omega$, while $k_F = (2{\cal M}_F
+3)^{1/2}\sqrt{m\omega/\hbar}\sim 4.3589\sqrt{m\omega/\hbar}$. The
quoted error bars take into account the minimization process itself
as well as effects of the initial conditions that could not be
erased during the thermalization process. It is important to point
out that the variational energy for the highest $1/k_Fa$ value
coincides with that obtained from a perturbative calculation
 using a contact interaction as can be verified from expressions
 obtained in Ref.~\cite{PRA04}.

\begin{figure}
\includegraphics[width=2.5in,height=2in,angle=0]{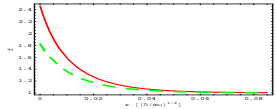}
\caption{(Color online) First choice Jastrow correlation function
for $b=0.03 \sqrt{\hbar/m \omega}$, dashed and solid lines
correspond to $1/k_Fa=-0.3873$ and $-15.8888$ respectively.
Distances are measured in units of $\sqrt{\hbar/m\omega}$.
}\label{healing}
\end{figure}

The optimal value of the variational parameter determines the
shape of the Jastrow correlation function. As an illustration, in
Fig.~\ref{healing} we plot the behavior of the optimal two
particle function $f_{ij}$ of Eq.~(\ref{Psivar}) for $b=0.03
\sqrt{\hbar/m \omega}$ and two different values of the scattering
length $1/k_Fa=-0.3873$ and $-15.8888$ respectively. We observe
that the distance at which $f_{ij}\sim 1$ is larger than the
potential range, this suggests long distance correlated pairs, as
expected for a BCS-like pairs. This result is also in good
agreement with previous findings reported in
\cite{carlson,bethe73} where the healing distance is used as a
variational parameter.

As  $a$ approaches the crossover region it is expected that the trial function
 Eq.~(\ref{eq:Jastrow}) will not describe properly the
 interatomic correlations:  pairing effects become essential so that the
 quantum numbers in the Slater determinants in $\Phi_{IFG}$ are
 not representative of the physical situation.

\subsubsection{Variational calculation on the BCS-BEC crossover region}
In a theory originally put forth by Eagles \cite{eagles} and later
by Leggett\cite{leggett}, it was proposed that a BCS wave function
of the form
\begin{equation}
\Psi_{\lambda_{EL}}= {\cal A}\left[ \phi(1_{\uparrow},1_{\downarrow})
\phi(2_{\uparrow},2_{\downarrow}) ...
\phi(N_{\uparrow},N_{\downarrow}) \right], \label{eq:BECwf}
\end{equation}
with ${\cal A}$ the antisymmetrizer operator that ensures the
correct properties under particle exchanges, was more generally
applicable than just to the weakly interacting limit \cite{eagles}:
a BCS-like wave function could eventually describe the ground state
from a Cooper pairing region to a BEC of composite bosons made up of
two fermions.

Following this point of view, here we propose a family of
single-parameter variational wave functions for the BCS-BEC
crossover regime, taking $\phi(i_{\uparrow},j_{\downarrow})$ as a
variational extrapolation of the ground state solution of the
trapped two body problem
\begin{equation}
\phi({\bf r}_{i\uparrow},{\bf r}_{j\downarrow})\cong
\varphi(r_{i,j})e^{-\lambda_{EL}|{\bf r}_{i\uparrow}+{\bf
r}_{j\downarrow}|^2/4}. \label{eq:becsingle}
\end{equation}
The variational parameter $\lambda_{EL}$ modulates the optimal shape
of the cloud. The wave function (\ref{eq:BECwf}) using the basis
(\ref{eq:becsingle}) guarantees that the Monte Carlo dynamics will
be guided by effects of both paired-particles relative ${\bf
r}_{ij}={\bf r}_{i\uparrow} - {\bf r}_{j\downarrow}$ and center of
mass ${\bf R}_{ij}=({\bf r}_{i\uparrow} + {\bf r}_{j\downarrow})/2$
vectors.

 It is worth to mention that at difference with previous
 calculations \cite{carlson,giorgini} here, by explicitly including
 easy interpretable inhomogeneous features in the wave function,
 we are able to explore the $trapped$ atoms as a whole as they evolve
 into the interacting regime.
 Besides, at difference with the mean field approach,
 no  optimal individual particle wave functions are searched, but
 the global effect of the interaction on the paired-particles wave
 function.

To estimate the energy in the BEC side, the algorithm described for
the BCS region is not useful because it depends on the explicit
structure of the wave function, written in a Jastrow-Salter form. In
order to set the variational energy  in a form suitable for Monte
Carlo estimations we exploit the two-body structure of the potential
and the primitive wave functions $\phi$. The antisymmetrized wave
function (\ref{eq:BECwf}) can be explicitly written as

\begin{equation}
\Psi_{\lambda_{EL}}=\sum_{\cal P} (-1)^{\cal P} \prod_{i=1}^{N}
\phi(i,{\cal P}(i)), \label{BEC}
\end{equation}
where the summation is taken over all possible permutations ${\cal
P}$ on set $\downarrow$, and $\phi(i,{\cal P}(i))$ are wave
functions having the form of Eq.~(\ref{eq:becsingle}) and argument
$({\bf r}_{i\uparrow},{\bf r}_{{\cal P}(i)\downarrow})$. We can
split the Hamiltonian of the system in a pair-like sum, using the
center of mass and relative coordinates of possible pairs as:

\begin{eqnarray}
H&=&\sum_{i}^{N} \Big[ \frac{{ p}^2_{i,{\cal P}_0(i)}}{2\mu}
+\frac{\mu}{2}\omega^2 r^2_{i,{\cal P}_0(i)}
\nonumber\\&+&V_{i,{\cal P}_0(i)}(r_{i,{\cal P}_0(i)})+
\frac{P^2_{i,{\cal P}_0(i)}}{2M}+\frac{M}{2}\omega^2 R^2_{i,{\cal
P}_0(i)}\Big] \nonumber \\&+& \sum_{i,j\neq {\cal P}_0(i)}
V(r_{i,j})
\end{eqnarray}
with ${\cal P}_0$ any given permutation.

Equation~(\ref{eq:becsingle}) let us write:
\begin{eqnarray}
H\Psi_{\lambda_{EL}}&=&\Big[N\epsilon_{0}+N\frac{3\hbar\omega\lambda_{EL}}{2}
\Big] \Psi_{\lambda_{EL}} \nonumber\\&+& (1-{\lambda^2_{EL}})
\sum_{i,{\cal P}}(-1)^{\cal P}\frac{M}{2}\omega^2 R^2_{i,{\cal
P}(i)}\prod_{l=1}^{N} \phi(l,{\cal P}(l))\nonumber
\\&+&  \sum_{\cal P}(-1)^{\cal P}\sum_{i,j\neq P(i)} V(r_{i,j}) \prod_{l=1}^{N}
\phi(l,{\cal P}(l))
\end{eqnarray}
with $\epsilon_0$ the ground state eigenvalue of the two
body-problem. To evaluate the last two terms via a Monte Carlo
simulation we proceed to complete the potential by adding and
subtracting the term used in the two-body solution, then:

\begin{eqnarray} H\Psi_{\lambda_{EL}} &=&\Big[
N\epsilon_{0}+N\frac{3\hbar\omega\lambda_{EL}}{2} + \sum_{i,j}
V(r_{i,j})\Big]\Psi_{\lambda_{EL}}  \nonumber\\ &+& \sum_{i,{\cal
P}}(-1)^{\cal P}\prod_{l\ne i}^{N} \phi(l,{\cal
P}(l))\cdot\Big[(1-{\lambda^2_{EL}})\frac{M}{2}\omega^2 R^2_{i,{\cal
P}(i)}\nonumber\\&-&V(r_{i,{\cal P}(i)}) \Big]\phi(i,{\cal
P}(i))\nonumber \\
\end{eqnarray}
which can also be written in terms of the minors $C_{i
\alpha}(\Psi_{\lambda_{EL}})$ associated to the $\Psi_{\lambda_{EL}}$:
\begin{equation}
 \Psi_{\lambda_{EL}}=\sum_{\alpha=1}^{N} C_{i \alpha}(\Psi_{\lambda_{EL}})\phi_{i,\alpha}
\end{equation}
where $\phi_{i,\alpha}$ represents any of the $i$-row wave
functions.
\begin{eqnarray} H\Psi_{\lambda_{EL}} &=&\Big[
N\epsilon_0+N\frac{3\hbar\omega{\lambda_{EL}}}{2} + \sum_{i,j}
V(r_{i,j})\Big]\Psi_{\lambda_{EL}}  \nonumber\\ &+& \sum_{i,\alpha}
C_{i,\alpha}\cdot\Big[(1-{\lambda^2_{EL}})\frac{M}{2}\omega^2
R^2_{i,\alpha}\nonumber-V(r_{i,\alpha}) \Big]\phi(i,\alpha)
\end{eqnarray}
This expression results quite convenient  for the simulations to
be performed, where one can also take advantage from the relation
between minors and the elements of the inverse of the
 transposed matrix \cite{ceperly},
\begin{equation}
{\bar \phi_{i,\alpha}}\equiv  (\phi^T)^{-1}_{i \alpha}=
\frac{C_{\alpha,i}(\Psi^T_{\lambda_{EL}})}{\Psi_{\lambda_{EL}}^T}=
\frac{C_{i,\alpha}(\Psi_{\lambda_{EL}})}{\Psi_{\lambda_{EL}}}.
\end{equation}
Thus, we can  write
\begin{eqnarray}
&&H\Psi_{\lambda_{EL}}=\Big[ N\epsilon_0+N\frac{3\hbar\omega{\lambda_{EL}}}{2} +
\sum_{i,j} V(r_{i,j})\nonumber\\
&+&\sum_{i,\alpha}^{N}\phi_{i,\alpha}\bar\phi_{i,\alpha}
\Big[(1-{\lambda^2_{EL}})\frac{M}{2}\omega^2
R^2_{i,\alpha}-V(r_{i,\alpha}) \Big] \Big]\Psi_{\lambda_{EL}}
\end{eqnarray}
As in the BCS calculation, we can sample the system using a
Metropolis-Monte Carlo algorithm and estimate the energy as a
function of the variational parameter.

 In Table I, we illustrate the results on the optimal variational
 parameter and different energy estimates for several scattering
 lengths. The strength of the potential was taken in the region
  around the first zero-resonance condition
  $(z_0/b)^2\hbar\omega =\tilde v_0$.
 The upper set of results were obtained using the Jastrow-Slater wave function
 Eqs.~(\ref{eq:Jastrow},\ref{Psivar}). All other results considered a
 wave function of the Eagles-Leggett form, Eq.~(\ref{eq:BECwf}), with the
 two-body functions Eq.~(\ref{eq:becsingle})
 taking  $\varphi(r_{i,j})$  as the approximate solutions of the two-body problem for a given
 scattering length, Eqs.(\ref{eq:form1}-\ref{eq:form2}).

 In the reported calculations using Eagles-Leggett wave functions,
 the range of the potential was taken as
 $b/2 =0.00375 \sqrt{\hbar/m \omega}$
 before the unitarity limit, $1/k_F a =0$, and $b/2 =0.0025 \sqrt{\hbar/m \omega}$
 for $k_F a \ge 0$. Actually, calculations were performed
 for several potential ranges $b/2$ all over the crossover. The $b/2$ ranges reported
 in this table are the shortest for which reliable numerical results
 were obtained.

 For $1/k_Fa < -0.45$ the
 variational energy $E/2N$ for the wave-function (\ref{eq:BECwf}-\ref{eq:becsingle})
 is higher than that obtained with the Jastrow-Slater
 trial wave-function,  while for $1/k_Fa > -0.45$ the situation is inverted and
 the $BCS$-wave function gives a lower upper bound for $E/2N$.
 A similar effect has been found in Ref.~\cite{giorgini}
 for the homogeneous gas. Beyond the unitarity region,
 $i.$ $e.$ for $a>0$, the contribution
 $\epsilon_0/2$ coming from  the trapped ground state two-body
 eigenvalue has been subtracted.
 For $4 <1/k_F a< 12$ an optimal value of ${\lambda_{EL}}\sim 1$ yielding a local
  minimum was found. However, for ${\lambda_{EL}} >1$ the corresponding mean value of the energy
 can be made arbitrarily small by considering ${\lambda_{EL}}$ large
 enough.   This variational instability is expected
 at the extreme BEC region for any attractive potential of
 {\it finite} range as discussed  previously in Ref.~\cite{pandharipandepra} for a
homogeneous gas. It could eventually be avoided by adding a
repulsive interaction at   distances much smaller than the range
$b/2$ as suggested in the original work by Leggett
 \cite{leggett}. Implementing this idea within our numerical approach is
 very difficult since we have already set $b\ll\sqrt{\hbar/m\omega}$.
Although a local minimum was found for $1/k_f a>4$, finite range
effects are expected to be significant on the reported data.

The wave function having ${\lambda_{EL}} \sim 1$ is expected for a
molecular gas when Pauli Blocking effects between the constituting
fermionic atoms are approximately compensated by the attractive
finite range interaction effects. If the atoms did not interact,
the corresponding energy per atom would be $E/2N=1.5\hbar\omega$
corresponding to an ideal gas of trapped Bose molecules. It is
also important to emphasize that, for a $contact$ interaction,
molecules formed by Fermi atoms are expected to have a weakly
repulsive interaction, with a molecule-molecule scattering length
given by $a_{mm}=0.6a$\cite{Shlyapnikov}.

\subsubsection{Virial relations}
  In  the fifth column of Table I, we also  report $\langle m\omega^2 R^2\rangle$,
  that is, twice the mean value of the trapping potential energy per particle,
  which is more feasible of experimental verification  \cite{regal} than the
  total energy $E$. Notice that in the crossover region  with
  $-0.1<1/k_Fa<1.4$, the energies in the third column are similar
  to $\langle m\omega^2 R^2\rangle$.

  At unitarity, $1/k_Fa=0$, a virial relation
  of the form $E/2N = \langle m\omega^2 R^2\rangle$
  is expected from previous  experimental and theoretical studies
  \cite{virial,wernercastin}. During the revision process
 of the present article, virial theorems for trapped interacting atoms
outside
 the unitarity limit have been established both at finite temperature \cite{tan,braaten}
 and at zero temperature \cite{thomas2,werner}. The latter
 considered several forms of the interaction potential. In particular for a
 contact interaction with a strength determined by the scattering length
 $a$, it has been shown that:
  \begin{eqnarray}
\frac{E}{2N} &=& \langle m\omega^2 R^2\rangle - \frac{1}{2}k_Fa\frac{\partial E/2N}{\partial k_Fa}\\
  &=& \langle m\omega^2 R^2\rangle + \frac{1}{2k_Fa}\frac{\partial
  E/2N}{\partial 1/k_Fa}.\label{eq:virial1}
\end{eqnarray}
For a contact interaction in the free space, the energy of the bound
state is $\epsilon_0^{c}=-\hbar^2/ma^2$. So that,  $ k_Fa\partial
\epsilon_0^{c}/\partial k_Fa =-2 \epsilon_0^{c}$ and, in the BEC
side of the crossover,
\begin{equation}
 \langle
m\omega^2 R^2\rangle^c_{virial}= (E/2N-\epsilon_0^c/2)-
\frac{1}{2}\Big[\frac{1}{k_Fa}\frac{\partial
  (E/2N-\epsilon_0^c/2)}{\partial 1/k_Fa}\Big].\label{eq:virialc}
\end{equation}
Although we have made all calculations with a finite range
potential we would like to evaluate how compatible our results are
with those arising from contact interaction predictions. Thus, in
the last column of Table I we report the
 mean value of twice the potential energy per particle associated to
the trap, evaluated numerically using Eq.~(\ref{eq:virial1}) in
the BCS side and the following expression in the BEC side:
\begin{equation} \langle m\omega^2 R^2\rangle_{virial}=
(E/2N-\epsilon_0^b/2)-
\frac{1}{2}\Big[\frac{1}{k_Fa}\frac{\partial
  (E/2N-\epsilon_0^b/2)}{\partial 1/k_Fa}\Big]\label{eq:virialb}
\end{equation}
with $\epsilon_0^{b}$ the two particle energy for a finite range
interaction state in otherwise free space. The  numerical
evaluation of the derivative was preceded by a numerical smoothing
of data. We observe that, although the calculations were performed
using a finite range potential, there is a reasonable agreement
between the trap energies and the virial expressions
Eqs.~(\ref{eq:virial1}) and (\ref{eq:virialb}) for
$-0.1<1/k_Fa<3.5$.

It was also  found that as $1/k_Fa \rightarrow 0^-$ and $1/k_Fa
\rightarrow 0^+$ the derivatives in
  Eq.~(\ref{eq:virial1}) and Eq.~(\ref{eq:virialb}) respectively attain a minimum.
  This minimum together with the small difference between the value of
$\epsilon_0/2$ and $\epsilon_0^b/2$ compared to $E/2N$ for
$0<1/k_Fa<1.5$, let us understand the observed similarities between
the third, fifth and sixth columns of Table I for $-0.1<1/k_Fa<1.4$.

On the BCS side of the crossover, for $-0.5<1/k_Fa<-0.2$, our wave
functions yield  a 15$\%$ higher trapping energy than the  virial
relation predicts. It is important to mention that, for these values
of $1/k_Fa$, the atomic cloud is quite extended and the evaluation
of both the mean energy $E/2N$ and the mean square radius $\langle
R^2\rangle$ requires
 special care of the statistics sampling. Improving the form of the variational
 wave function in this region, could diminish the discrepancy with the
 virial relation. Notice that, in the language of BCS theory, these
 region delimits
 the transition of the atomic cloud from a normal to a superfluid state.

In the region $1/k_F a>4$ there is also a discrepancy with the
contact virial relation; it could be due to finite range potential
effects as expected from the variational instability reported above.
In fact, for these scattering lengths the difference between atoms
interacting through a contact and a finite range potential is
already evident by comparing
 their corresponding two-body ground state energies $\epsilon_0$.
  These energies are in general similar but, as expected, the bigger differences
 appear for the deeply bound two-body states,
 $a\rightarrow 0^+$. For instance,
 at the bottom of Table I, $\epsilon_0=-756.6\hbar\omega$ in contrast to the solution of
 Eq.~(\ref{eq:wilkens}) which yields -578.1$\hbar\omega$;
 for the other states the difference is less than $10\%$ up to $1/k_Fa <2.5$
 and around 15$\%$ for the remaining reported data.

\begin{table}[t]
\begin{tabular}{|c|c|c|c|c|c|}
\hline
 $1/k_F a$ &$V_0\lambda^{opt}_{J1}$ & $E/2N$ &
 $\epsilon_0$&$<m\omega^2 R^2>$&$<m\omega^2 R^2>_{virial}$\\
           &    $\pm 0.05$                  & [$\hbar
           \omega$]&$[\hbar\omega]$& $[\hbar\omega]$& $[\hbar\omega]$\\
\hline -15.88895 & -0.6 &  7.450$\pm$ 0.004&1.49&7.5$\footnote {The
evaluation of mean radii for extended clouds requires
special care of the statistics.}\pm$0.2&7.418$\pm$ 0.006\\
-9.60883 & -0.6 & 7.425$\pm$ 0.005 & 1.48&7.5$^a\pm$0.2&7.39$\pm$ 0.0075\\
-4.26632 & -0.6 & 7.374$\pm$0.009&1.46& 7.53$^a \pm$0.2&7.31$\pm$ 0.03\\
-2.02479 &-0.9 &7.345$\pm$0.012&1.42&7.53$^a\pm$0.2&7.25$\pm$0.03\\
\hline \hline$1/k_F a$ &$\lambda^{opt}_{EL}\hbar/m\omega$ & $E/2N$ &
 $\epsilon_0$&$<m\omega^2 R^2>$&$<m\omega^2 R^2>_{virial}$\\
           &            $\pm 0.005$         & [$\hbar
           \omega$]&$[\hbar\omega]$& $[\hbar\omega]$& $[\hbar\omega]$\\
           \hline
-0.43893 & 0.142 & 7.07$\pm$0.06&1.15&7.72$^a\pm$0.2&6.88$\pm$0.09\\
-0.22418 & 0.134 & 6.65$\pm$0.06&0.99 &7.57$^a\pm$0.2&6.42$\pm$ 0.09\\
-0.10071 & 0.160 &6.37$\pm$0.06 &0.79&6.58$\pm$0.2&6.21$\pm$0.09\\
-0.03745 & 0.182&6.10$\pm$0.06 &0.72&5.64$\pm$0.2&5.88$\pm$ 0.09\\
 0 &0.186 & 5.25$\pm$0.08&0.50&5.32$\pm$0.2&5.25$\pm$0.12\\
 \hline
 \hline
 $1/k_F a$ &$\lambda^{opt}_{EL}\hbar/(m\omega)$ & $(E/2N)-\epsilon_0/2$ &
 $-\epsilon_0$&$<m\omega^2 R^2>$&$<m\omega^2 R^2>_{virial}$\\
           & $\pm 0.005$                     & [$\hbar
           \omega$]&$[\hbar\omega]$& $[\hbar\omega]$& $[\hbar\omega]$\\
 \hline
 0.13959& 0.190& 4.78$\pm$0.07 &0.21&4.92$\pm$0.2&4.85$\pm$0.11\\
 0.34876& 0.191 & 4.18$\pm$ 0.07 &2.45&4.50$\pm$0.2&4.52$\pm$0.11\\
0.69684 & 0.256 & 3.57 $\pm$0.07&9.95&3.75$\pm$0.2&3.95$\pm$0.2\\
1.04427 & 0.270 & 3.35$\pm$0.06&22.65&3.51$\pm$0.2&3.85$\pm$0.2\\
1.39107 &0.380&3.14$\pm$0.08&40.76 &2.99$\pm$0.2&3.69$\pm$0.2\\
2.08293 & 0.665&2.60$\pm$0.08& 93.96&2.95$\pm$0.2&3.38$\pm$0.2\\
2.77266 &0.90&2.0$\pm$0.1&171.157&2.89$\pm$0.2&3.09$\pm$0.2\\
 3.46055 & 0.94 &1.38$\pm$0.15&274.99 &2.59$\pm$0.25&2.45$\pm$0.23\\
 4.1469& 0.99 & 1.1$\pm$0.15&404.64& 2.59$\pm$0.25&1.71$\pm$0.23\\
 5.51634 &0.99& 0.95$\pm$0.15&756.643 &2.59$\pm$0.40&1.19$\pm$0.23\\
\hline
\end{tabular}
 \caption{ Optimal variational
parameter $\lambda$,
 energy per particle, two-body ground $\epsilon_0$ energies,
 mean value of the trap potential energy per particle
  $<m\omega^2 R^2>$ from Monte Carlo calculations and $<m\omega^2 R^2>_{virial}$ evaluated
 using the virial relation given by Eq.~(\ref{eq:virial1},\ref{eq:virialb}).  All of them were
 calculated as a function of $1/k_Fa$ considering $2N=330$ particles. The upper set of
 results used the Jastrow-Slater wave function
 Eqs.~(\ref{eq:Jastrow},\ref{Psivar}). All other results considered a
 wave function of the Eagles-Leggett form, Eq.~(\ref{eq:BECwf}), with the
 two-body functions  $\varphi(r_{i,j})$ taken
 as the approximate solutions of the two-body problem for a given
 scattering length.}
\end{table}

\subsubsection{Unitarity}

At unitarity, $\vert a\vert\rightarrow \infty$, we have estimated
the energy using  a variational wave function of the form
Eq.~(\ref{BEC}).  The numerical results were presented and broadly
discussed in Ref.~\cite{rapidA} in connection with the universality
hypothesis. Notice that, in Table I, the numerical errors at
unitarity are  slightly larger than those obtained for nearby
scattering lengths. The reason can be traced back to the qualitative
difference between the two-body wave function $u_{ij}=r\varphi_{ij}$
determined by Eq.~(\ref{eq:form2}). The high delocalization of the
unitarity paired-atoms wave function  makes more difficult the
evaluation of the energy expectation value at this limit.
 From these results, an upper bound to the universal parameter $\beta$, defined
by $E_U=E_{IFG} \sqrt{1+\beta}$, is found to be $\beta =
-0.51\pm0.01$.

\begin{figure}
\includegraphics[width=3in]{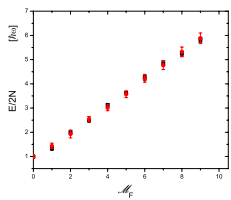}
\caption{(Color online) $s$-ground state variational energy per atom
$E/2N$ and $\langle m\omega^2R^2\rangle$ for trapped particles at
unitarity as a function of the Fermi number ${\cal M}_F$ for closed
shells. Energy is measured in units of $\hbar\omega$.}
\label{ground}
\end{figure}

In Fig.~\ref{ground}, we show the mean value $<m\omega^2 R^2({\cal
M_F})>$ together with $E({\cal M_F})$  for closed shells with ${\cal
M_F}\le 9$ corresponding to $N=$ 4, 10, 20,35, 56, 84, 120, 165 and
220 particles. No significant discrepancy among these mean values is
observed and the virial relation is thus verified.
 As reported in Ref.\cite{rapidA}, a linear relationship between
 the energy per particle at unitarity and
the shell number ${\cal M_F}$ is also found:
\begin{equation}
E_U/2N \sim (0.53\pm 0.01)({\cal M}_F + (1.95\pm 0.06))\hbar\omega,
\end{equation}
when this expression is compared with the ideal Fermi gas result,
Eq.~(\ref{dtfg}), one obtains an upper bound for the universal
parameter $\beta=-0.50^{(+0.02)}_{(-0.04)}$.

 \subsection{ Densities and correlations}

The information encoded in the single-particle and the two-particle
correlation functions is important since those functions reflect the
quantum mechanical nature of the particles and their collective
behavior, driven by the interaction and trapping potentials.

In the following, we illustrate these correlation functions for
$N=165$ using the optimized wave functions in the BCS  ($a<0$), the
unitarity ($a \to \infty$) and the molecular ($a>0$) regimes. We
have chosen examples in the crossover with $\vert 1/k_Fa\vert <1 $
due to its expected independence on the details of the calculation.

For completeness, let us recall that the single-particle correlation
function, that is, the density profile as a function of the distance
to the center of the harmonic trap, has already been illustrated in
Fig.~2 of Ref.~\cite{rapidA}. There, we saw that the trap effect is
reflected by decreasing the particle density until vanishing around
the Fermi radius, i.e., the inhomogeneous environment
 created by the harmonic confinement affects all the regimes
 as it is already evident
 for an ideal Fermi gas in the Thomas-Fermi approximation
 \cite{Butts}. The shape of the BCS density profile
 is similar to the one corresponding to the
  ideal gas but with a different mean radius. The density increases at
the center while decreases as it goes to the edge of the trap. These
deviations can be attributed to
  the optimal value of the variational parameter  which captures
  the interaction and correlation
 effects in the many-body system. This kind of shape prevails up
 to the unitarity limit.
The major differences in the particle density for each regime occur
around the center of the trap, particularly for the BEC regime where
most of the paired atoms are located near the origin.

\begin{figure}
\includegraphics[width=2.5in,height=2in,angle=0]{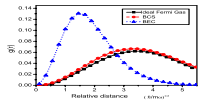}
\caption{(Color online) Normalized correlation function $g(r)$ for
particles in the same hyperfine state as a function of their
relative distance. Square, circle, and triangle symbols correspond
to the ideal Fermi gas, BCS ($1/k_Fa=-0.224$) and BEC
($1/k_Fa=0.697$) respectively. Distances are measured in units of
$\sqrt{\hbar/m\omega}$.}\label{pauli}
\end{figure}

 In order to exhibit the quantum behavior of
the fermionic atoms, the two-particle correlation function for
particles in the same hyperfine state, $g(r)$, was computed. The
calculations involved  finding the fraction of atoms in the same
hyperfine state within a relative distance ($r$, $r$ + d$r$), as
generated by the Monte Carlo sampling, irrespective of the center of
mass position; $g(r)$ was normalized dividing by $N(N-1)/2$ to
account for the combinatorial of the atoms. Figure \ref{pauli}
illustrates the resulting correlation functions. The Jastrow-Slater
wave function in the limit of an ideal Fermi gas ($\lambda_{J1}
=0$),
 exhibits the Pauli blocking arising from the fermionic
 nature of the atoms.
 The BCS  trial wave function  shows an slightly
 diminished Pauli blocking for short distances. In the
 molecular side, it is observed that particles in the same hyperfine state
 can be found around the same region. Although at
 the  deep BEC regime Pauli blocking  still
  inhibits the presence of atoms in the same hyperfine state, the
  radius at which it is evident becomes very short. As a consequence,
   if an exclusively attractive
  interacting potential is considered and it is large enough, Pauli blocking
  is not able
  to avoid a variational collapse as discussed above.
  All of these correlations  decrease
  for long relative distances as a consequence of the presence of the trap.

The two-particle correlation function for atoms in different
hyperfine states was computed in  Ref.\cite{rapidA}. There (Fig.~3)
its behavior in the BEC regime is compared with respect to the ideal
regime, as a function of the relative distance $r_{i,j}= |{\bf
r}_{i\uparrow} - {\bf r}_{j\downarrow}|$ among them. It was
evaluated in a similar way to $g(r)$, taking care of the proper
normalization factor ($N^2$) and keeping the information of the
center of mass position of the pairs. Molecule formation was
indicated by the increase in the correlation for very short
distances, $r_{i,j}\ll\sqrt{\hbar/m\omega}$. Most molecules are
formed for $R_{cm}<1.09\sqrt{\hbar/m\omega}$. An enhancement of the
probability of finding pairs of particles separated at relative
distances of the order of $r_{i,j}\sim \sqrt{\hbar/m\omega}$
indicated molecular condensation effects.

\begin{figure}
\includegraphics[width=2.5in,height=2in,angle=0]{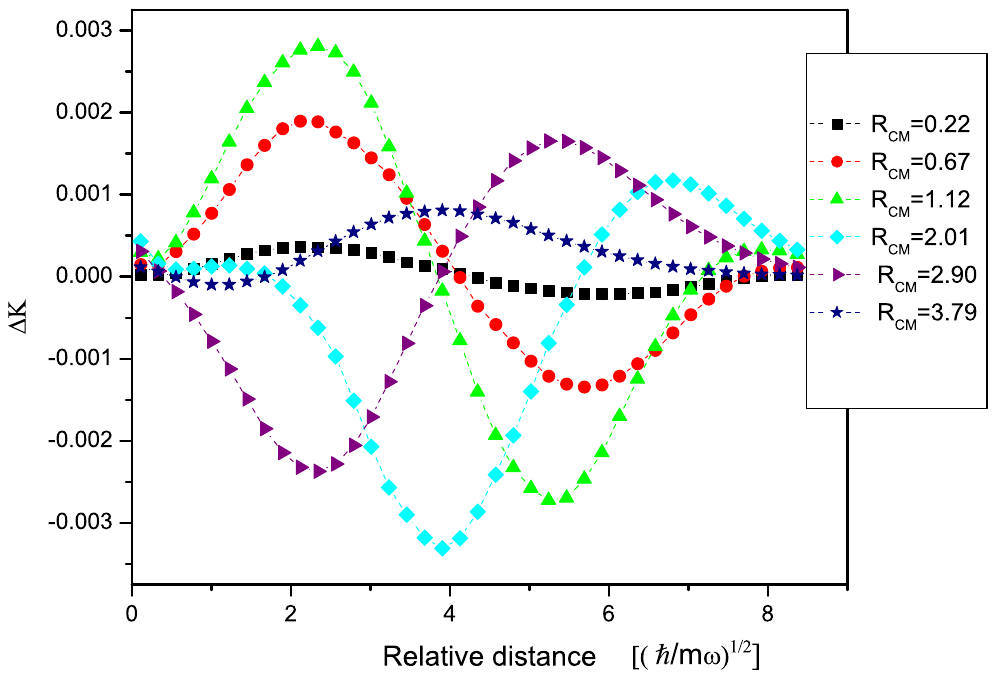}
\caption{(Color online) Probability difference $\Delta
K(r_{i,j},R_{cm})$ that two particles with different spin are
found separated a distance $r_{i,j}$ in the BCS and ideal regimes.
Each curve in this figure correspond to a spherical radius
$R_{cm}$ measured from the center of the trap. Calculations are
performed at $1/k_Fa=-0.224$. Distances are measured in units of
$\sqrt{\hbar/m\omega}$.}\label{bcs-free}
\end{figure}

Here, in Fig.~\ref{bcs-free} we illustrate the differences between
the two-particle correlation functions of atoms in different
hyperfine states, $\Delta K(r_{i,j},R_{cm})$  for the ideal and
BCS regimes. As in Fig.~3 of Ref.~\cite{rapidA}, it shows results
for a set of radius $R_{cm}$ measured from the center of the trap.
We observe that, although not zero, the difference is very small
(see the abscissa scale) compared to the result for the BEC
regime, in addition strong oscillations in $r_{i,j}$ are seen for
all $R_{cm}$.

\section{CONCLUSIONS}

We have studied an interacting two-component Fermi gas confined in
an isotropic harmonic potential in three dimensions. To be specific,
we investigated the transition from a Bardeen-Cooper-Schrieffer
state to a Bose-Einstein condensate  at zero temperature for a
system composed of $N=165$ particles of equal mass in each
spin-state. The interaction between particles of different spin was
considered to be an attractive potential with very short range
interaction; under such conditions it is expected that the many-body
ground state depends just on the product of the scattering length
$a$ and the Fermi wave number $k_F$. The BCS-BEC transition was
followed as a function of $k_Fa$.

To model the gas, we proposed a family of many-body trial wave
functions for the BCS ($a<0$) and the BEC ($a>0$) sides. For small
negative values of $a$ we described the atomic gas by a
Jastrow-Slater wave function. While, for other values of $a$,
following  Eagles and Leggett proposal, a wave function written as
the antisymmetric product of two-particle states was used. The
two-body basis was formed by analytical compact  functions that
contain collision and trapping effects.  For a given interaction
range, using variational Quantum Monte Carlo simulations, we found
the variational parameter $\lambda^{opt}$ that minimizes the energy
per particle of the whole system.  Efficient algorithms to estimate
the energy expectation value, exploiting properties of the
antisymmetrized many-body functions, were elaborated to perform
calculations for such a large number of particles. After considering
several values of the range of the potential and studying the
stability of the results, we reported the numerical data
corresponding to the lowest value of the potential range that gave
reliable numerical results. The corresponding optimal variational
wave functions lead to predictions for the main properties of the
trapped system like energies, mean squared radii, one and two-point
correlation functions.

 The system energy was
computed all along the crossover, and at unitarity an upper bound to
the universal parameter $\beta$ is found to be $\beta =
-0.51\pm0.01$. This result is compatible with the result reported in
Ref.~\cite{rapidA} where $\beta_{fit} = -0.50^{-0.02}_{+0.04}$ was
found by comparing $E_U$ and $E_{IFG}$ for  ${\cal M_F}\le 9$. Those
calculations indicate that the universal hypothesis yields results
consistent with theoretical calculations even for a small $N$. So
that, for zero temperature, the energy of a balanced mixture of
interacting trapped fermions has the form $E_U \sim 1/2({\cal
M_F}+2)2N\hbar\omega$ similar to the ideal Fermi gas equation
$E_{IFG}=3/4({\cal M_F}+2)2N\hbar\omega$. In addition it was shown
that not only at unitarity but also over the crossover region
$-0.1<1/k_Fa<1.4$ the mean value of the atomic gas squared radius
can be used to give a rough estimate of the energy per particle,
since $<m\omega^2R^2>\sim E/2N -\bar\epsilon_0/2$, where
$\bar\epsilon_0$ is the two-body ground state energy $\epsilon_0$
for trapped fermions for $a>0$ and zero for $a\le 0$ and $\vert
a\vert\rightarrow \infty$ .

In agreement with previous results \cite{rapidA}, the energy
function  $E(1/k_Fa)$ along the crossover follows a curve that
properly normalized is independent of ${\cal M}_F$. By evaluating
its numerical derivative a minimum was found at unitarity . Besides,
this function was shown to satisfy a virial relation  in the
interval $-0.1<1/k_Fa<3.4$. This relation was based on virial
theorems for trapped atoms developed by other groups in the last few
months \cite{tan,braaten,thomas2,werner}.

The starting point of the crossover from the BCS side could be
regarded as the value of $1/k_Fa$ for which the antisymmetric
product of two-particle states gives a lower expectation value of
the energy with respect to the Jastrow-Slater wave function.
According to our calculations this already occurs at $1/k_Fa \sim
-0.45$. This value is similar to that at which a variational
calculation based on scaled antisymmetric product of harmonic
oscillator wave functions can not be applied since no minima exists
\cite{PRA04}. Although the Eagles-Leggett wave function, built using
the solutions of the two-body problem, gives a better description of
the system than its Jastrow-Slater analog, it yields a  mean radius
for the atomic cloud 15\% larger than that predicted by the virial
relations. Thus, it would be important to work out an improved trial
wave function to describe this transition zone.

In the extreme BEC region there is a variational instability which
arises from the usage of $finite$ range attractive potentials
between the fermions. In our calculations the extreme BEC region
starts when the variational parameter $\lambda_{EL}$ yields a local
minimum energy for $\lambda^{opt}_{EL}>1$. In such a case the effect
of Pauli blocking is supersede by the very strong short range
attractive potential. For $12>1/k_Fa>3.5$ and
$b/2=0.0025\sqrt{\hbar/m\omega}$ a local minimum was found with
$\lambda_{EL}\sim1$. This value of $\lambda_{EL}$ corresponds to a
many-body wave function of an ideal Bose gas of trapped molecules.
This function does not satisfy the virial relation for a contact
interaction so that even though $b/2<<\sqrt{\hbar/m\omega}$, finite
range effects are not negligible for those values of $k_Fa$.

Finally, we calculated the one-particle and the two-particle
correlation functions for the BCS and BEC regimes and for the
unitary limit. The results show that the  correlation length between
pairs can be much larger than the interaction potential range.  As
expected, the inhomogeneous environment resulting from the harmonic
confinement affects all the regimes. We  observe that in the BCS
regime, the paired atoms have a large correlation length
particularly for $0.6<R_{cm}<1.2 \sqrt{\hbar/m\omega}$. Pauli
blocking effects were also sensible to trapping and interaction
strength. Thus, we conclude that the approximate analytical wave
function used to describe the trapped interacting gas gives a good
compact representation of the system through the crossover region.

{\bf Acknowledgments} This work was partially supported by Conacyt
M\'exico, under grant 41048-A1 and DGAPA-UNAM contract PAPIIT
IN117406-2.


\begin{thebibliography}{99}

\bibitem{D.Jin} B. DeMarco and D. S. Jin, Science {\bf 285}, 1703 (1999).

\bibitem{experiments} S. Jochim, M. Bartenstein, A. Altmeyer, G. Hendl,
S. Riedl, C. Chin, J. H. Denschlag, and R. Grimm, Science {\bf 302},
2101 (2003); M. Greiner, C. A. Regal, and D. S. Jin, Nature {\bf
426}, 537 (2003); M. W. Zwierlein, C. A. Stan, C. H. Schunck, S. M.
F. Raupach, S. Gupta, Z. Hadzibabic, and W. Ketterle, Phys. Rev.
Lett. {\bf 91}, 250401 (2003); K. E. Strecker, G. B. Partridge, and
R. G. Hulet, Phys. Rev. Lett. {\bf 91}, 080406 (2003).

\bibitem{thomas} K. M. O'Hara, S. L. Hemmer, M. E. Gehm, S. R. Granade,
 J. E. Thomas, Science {\bf 298}, 2179(2002); M. E. Gehm, S. L.  Hemmer,
 S. R. Granade, K. M. O'Hara, and J. E. Thomas, Phys. Rev. A {\bf
68}, 011401(R) (2003).

\bibitem{bartenstein} M. Bartenstein, A. Altmeyer, S. Riedl, S. Jochim,
C.  Chin, J. H. Denschlag, and R. Grimm, Phys. Rev. Lett. {\bf 92},
120401 (2004).

\bibitem{salomon} T. Bourdel, J. Cubizolles, L. Khaykovich,
K. M. F. Magalhaes, S. J. J. M. F. Kokkelmans, G. V. Shlyapnikov,
and C. Salomon, Phys. Rev. Lett. {\bf 91}, 020402 (2003).

\bibitem{bourdel} T. Bourdel, L. Khaykovich, J. Cubizolles, J. Zhang,
F. Chevy, M. Teichmann, L. Tarruell, S.J.J.M.F. Kokkelmans, and C.
Salomon, Phys. Rev. Lett. {\bf 93}, 050401 (2004).

\bibitem{heiselberg} H. Heiselberg,  Phys. Rev. A {\bf 63},
043606 (2001).

\bibitem{ho}T.-L. Ho,  Phys. Rev. Lett. {\bf 92}, 090402 (2004).

\bibitem{wernercastin} F. Werner and Y. Castin, Phys. Rev. A {\bf
74}, 053604 (2006).

\bibitem{field-theory} W. Vincent Liu, Phys. Rev. Lett, {\bf 96}, 080401 (2006).

\bibitem{carlson}J. Carlson, S. Y. Chang, V. R. Pandharipande,
and K. E. Schmidt, Phys. Rev. Lett. {\bf 91}, 050401 (2003).

\bibitem{pandharipandepra} S. Y. Chang, V. R. Pandharipande, J. Carlson, and K. E.
Schmidt, Phys. Rev. A {\bf 70}, 043602 (2004).

\bibitem{giorgini}G. E. Astrakharchik, J. Boronat, J. Casulleras,
and S. Giorgini, Phys. Rev. Lett. {\bf 93}, 200404 (2004).

\bibitem{Dlee} D. Lee, Phys. Rev. B {\bf 73}, 115112 (2006).


\bibitem{chang07} S. Y. Chang and G. F. Bertsch, Phys. Rev. A {\bf 76},
 021603(R) (2007).

\bibitem{BlumeD07} D. Blume, J. von Stecher, C. H. Greene, Phys. Rev.
Lett. {\bf 99}, 233201 (2007).

\bibitem{BlumeD08} J. von Stecher and C. H. Greene, and D. Blume, Phys. Rev. A {\bf 77},
043619 (2008).



\bibitem{kinast} J. Kinast, A. Turlapov, J. E. Thomas, Q. J. Chen, J. Stajic,
and K.  Levin, Science {\bf 307}, 1296 (2005).

\bibitem{hulet}
G. B. Partridge, W. Li, R. I. Kamar, Y. an Liao, and R. G. Hulet,
Science {\bf 311}, 503 (2005).

\bibitem{regal} J. T. Stewart, J. P. Gaebler, C. A. Regal, and D. S. Jin,
Phys. Rev. Lett. {\bf 97}, 220406 (2006).

\bibitem{rapidA} R. J\'auregui,  R. Paredes and G. Toledo S\'anchez, Phys. Rev. A
{\bf 76}, 011604(2007).


\bibitem{bolda}E. L. Bolda, E. Tiesinga, and P. S. Julienne,
Phys. Rev. A {\bf 66}, 013403 (2002)

\bibitem{rarita} W. Rarita and R. D. Present,  Phys. Rev.
{\bf 51}, 788 (1937).

\bibitem{wilkens}
T. Busch, B.G. Englert, K. Rza\c{a}\.zewski, and M. Wilkens,
Foundations of Phys. {\bf 28}, 549 (1998).

\bibitem{Daniel} D. Schiff and L. Verlet, Phys. Rev. {\bf 160}, 208 (1967).

\bibitem{bethe73}  V. R. Pandharipande and H. A. Bethe, Phys. Rev.
C {\bf 7}, 1312 (1973).

\bibitem{pandharipande77} V. R. Pandharipande and K.E. Schmidt,
Phys. Rev. A {\bf 15}, 2486 (1977).

\bibitem{nuclear} C. J. Horowitz, E.~J.~Moniz and J.~W.~Negele,
Phys. Rev. D {\bf 31}, 1689 (1985); C. J. Horowitz
and J.~Piekarewicz, Nucl. Phys. A {\bf 536}, 669 (1992).

\bibitem{toledoPRC02}   G. Toledo S\'anchez, and J. Piekarewicz, Phys.~
                   Rev.~ C{\bf 65}, 045208 (2002).

\bibitem{ceperly} D. Ceperley, G.V. Chester and M.H. Kalos,
Phys. Rev. B {\bf 16}, 3081 (1977).

\bibitem{PRA04} R. J\'auregui, R. Paredes and G. Toledo S\'anchez,
 Phys. Rev. A. {\bf 69} 013606 (2004).

\bibitem{Ko86}     Steven E. Koonin, {\it ``Computational Physics''}
(Benjamin Cummings, Menlo Park, 1986).

\bibitem{Me53}     N. Metropolis, A. Rosenbluth, M. Rosenbluth, A. Teller,
  and E. Teller, J. Chem Phys. {\bf 21}, 1087 (1953).

\bibitem{baker} G. A. Baker, Phys. Rev. C {\bf 60}, 054311
(1999).

\bibitem{Butts} D.A. Butts and D.S. Rokhsar, Phys. Rev. A {\bf 55}, 4346 (1997).

\bibitem{eagles}D. M. Eagles, Phys. Rev. {\bf 186}, 456 (1969).

\bibitem{leggett} A. J. Leggett, J. Phys. (Paris) {\bf 41}, C7 (1980).

\bibitem{Shlyapnikov} D.S. Petrov, C. Salomon, and G.V. Shlyapnikov,
Phys. Rev. Lett. {\bf 93}, 090404 (2004).

\bibitem{virial} J. E. Thomas, J. Kinast, and A. Turlapov, Phys.
Rev. Lett. {\bf 95}, 120402 (2005).



\bibitem{tan} S. Tan, e-print arXiv:0803.0841.

\bibitem{braaten} E. Braaten and L. Platter, Phys. Rev. Lett. {\bf
100}, 205301 (2008).

\bibitem{thomas2} J. E. Thomas, Phys. Rev. A {\bf 78}, 013630 (2008).

\bibitem{werner}F. Werner, Phys. Rev. A  {\bf 78}, 025601 (2008).



\end{thebibliography}
\end{document}